# Synchronization Analysis in Physical Layer Network Coding

Shengli Zhang, Soung-Chang Liew, and Hui Wang

*Abstract*—**Physical-layer Network Coding (PNC) makes use of the additive nature of the electromagnetic (EM) waves to apply network coding arithmetic at the physical layer. With PNC, the destructive effect of interference in wireless networks is eliminated and the capacity of networks can be boosted significantly. This paper addresses a key outstanding issue in PNC: synchronization among transmitting nodes. We first investigate the impact of imperfect synchronization (i.e., finite synchronization errors) in a 3-node network. It is shown that with QPSK modulation, PNC still yields significantly higher capacity than straightforward network coding when there are synchronization errors. Significantly, this remains to be so even in the extreme case when synchronization is not performed at all. Moving beyond a 3-node network, we propose and investigate a synchronization scheme for PNC in a general chain network. At last, numerical simulation verifies that PNC is robust to synchronization errors. In particular, for the mutual information performance, there is about 0.5dB loss without time synchronization and there is at most 2dB loss without phase synchronization.**

# I. INTRODUCTION

One of the biggest challenges in wireless communication is how to deal with the interference at the receiver when signals from multiple sources arrive simultaneously. In the radio channel of the physical layer of wireless networks, data are transmitted through electromagnetic (EM) waves in a broadcast manner. The interference between these EM waves causes the data to be scrambled. While interference has a negative effect on wireless networks in general, its detrimental effect on the throughput of *multi-hop* ad hoc networks is particularly noticeable [1, 2,3].

Most communication system designs try to either reduce or avoid interference (e.g., through receiver design or transmission scheduling [1]). However, instead of treating interference as a nuisance to be avoided, we can actually embrace interference to improve throughput performance. To do so in a multi-hop network, we proposed Physical-layer Network Coding (PNC) in [4] to create an apparatus similar to that of network coding, but which performs network coding arithmetic at the lower physical layer using the additive property of EM signal reception. Through a proper modulation-and-demodulation technique at relay nodes, addition of EM signals can be mapped to $GF(2^n)$ addition of digital bit streams, so that the interference becomes part of the arithmetic operation in network coding.

Two levels of synchronization between the two end-nodes were assumed in PNC in [4], namely symbol-level time synchronization, and carrier- frequency/phase synchronization. In this paper, we first investigate the impact of imperfect synchronization (i.e., finite synchronization errors) on PNC in a 3-node network with QPSK modulation. It is shown that PNC still yields significantly higher capacity than straightforward network coding when there are synchronization errors. Significantly, this remains to be so even in the extreme case when synchronization is not performed at all. Moving beyond the 3-node linear network, we propose a synchronization scheme for PNC in the *N*-node linear network. The *N*-node network can be decomposed into a chain of 3-node PNC units for synchronization purposes. We argue that if channel coding is applied on each of the 3-node PNC units, then the performance in terms of the end-to-end capacity will be the same for the *N*-node network and the 3-node network.

*Related Work*:

Synchronization has long been an active research problem in wireless networks. Here, we here review prior work on synchronization relevant to the 3-node PNC case. First, symbol time and carrier-frequency synchronizations, which are needed in PNC, have been actively investigated by researchers in the fields of OFDMA, wireless-sensor network, and cooperative transmission. In particular, methods for joint estimation of carrier-frequency errors, symbol timing error and channel response have been proposed for OFDMA networks [5, 6], and methods for symbol synchronization have been proposed in wireless sensor networks [7, 8] All these methods can be borrowed to finish the symbol time and frequency synchronization in PNC. Besides, PNC needs the critical carrier phase synchronization (which also implies a very accurate carrier frequency synchronization), which has recently been studied in the field of coherent cooperation (distributed beam forming ) to synchronize the separately distributed nodes. Among all these schemes, the most direct scheme uses the beacons bounded between the destination (relay) node and the source (end) nodes to estimate the relative phase offsets, so that each source compensated its phase according to this offset before transmitting [9]. To reduce the high feedback rate required in [9], a one bit feed back synchronization scheme is proposed in [10] and showed good performance with experiments. Removing the iterative information exchanging between the destination and the source nodes, some open loop algorithms are proposed. For example, positive results have been obtained in [11] with a master-slave architecture to prove the feasibility of the distributed beam forming technique. In [12, 13], another open loop synchronization scheme, round trip synchronization, were proposed and discussed where a beacon is used to measure round trip phase delays among the transmitters and the destination. We can use the ideas in these schemes to synchronize the phase of the two end nodes in the three-node PNC schemes. Besides the proposed synchronization algorithms, people also analyzed the affect of the synchronization errors in the coherent cooperate transmissions. For example, a more realistic collaborative communication system that includes the influence of AWGN and phase error on the signal transmission is analyzed in [14]. However, the analysis in [14] is different from the analysis in our paper since the source nodes in [14] transmit the same signal and the source nodes in our paper transmit different signals.

The rest of this paper is organized as follows. Section II presents the system model and illustrates the basic idea of PNC with a linear 3-node network under the assumption of perfect synchronization. Section III analyzes the performance penalty of non-perfect synchronization on PNC. Section IV proposes a strategy to extend 3-node synchronization to $N$-node synchronization in a long PNC chain. Section V studies the performance penalty of non-perfect synchronization by numerical simulation, and section VI concludes the paper.

II. System Model and Illustrating Example

## A. System Model:

In this paper, we focus on two-way relay channels (TWRC). A typical TWRC is the three node two way relay channel as shown in Fig. 1. $N_1$ (Node 1) and $N_3$ (Node 3) are nodes that exchange information, but they are out of each other's transmission range. $N_2$ (Node 2) is the relay node between them. This system model has found applications in many scenarios. In satellite communication, the satellite serves as a relay to facilitate information exchange between two mobile stations on the earth. In wireless mesh networks, wireless nodes may also relay information between its neighbors.

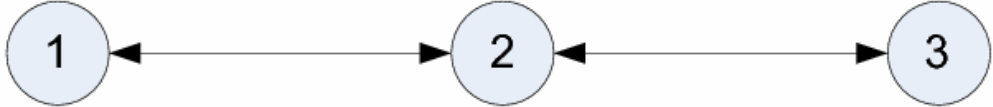

Figure 1. A 3-node linear network

We consider frame-based communication in which a time slot is defined as the time required for the transmission of one fixed-size frame. In this paper, a frame is denoted by a capital letter and symbols within a frame are denoted by the corresponding small letter. Each node is equipped with an omni-directional antenna, and the channel is half duplex so that transmission and reception at a particular node must occur in different time slots. We assume that QPSK modulation is employed at all the nodes.

## B. Physical-layer Network Coding Scheme

Before introducing the PNC transmission scheme, we first describe the traditional transmission scheduling scheme and the "straightforward" network-coding scheme for mutual exchange of a frame in the 3-node network [15, 16].

The traditional transmission schedule is given in Fig. 2. Let $S_i$ denote the frame initiated by $N_i$. $N_1$ first sends $S_1$ to $N_2$, and then $N_2$ relays $S_1$ to $N_3$. After that, $N_3$ sends $S_3$ in the reverse direction. A total of four time slots are needed for the exchange of two frames in opposite directions.

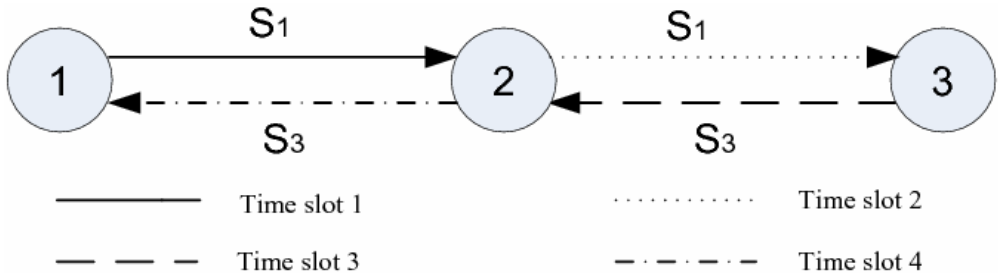

Figure 2. Traditional scheduling scheme

Ref. [15] and [16] outline the straightforward way of applying network coding in the 3-node wireless network. Fig. 3 illustrates the idea. First, $N_1$ sends $S_1$ to $N_2$ and then $N_3$ sends frame $S_3$ to $N_2$. After receiving $S_1$ and $S_3$, $N_2$ encodes them to obtain the frame $S_2 = S_1 \oplus S_3$, where $\oplus$ denotes bitwise exclusive OR operation being applied over the entire frames of $S_1$ and $S_3$. $N_2$ then broadcasts $S_2$ to both $N_1$ and $N_3$. When $N_1$ ($N_2$) receives $S_2$, it extracts $S_3$ ($S_1$) from $S_2$ using the local information $S_1$ ($S_3$). A total of three time slots are needed, for a throughput improvement of 33% over the traditional transmission scheduling scheme.

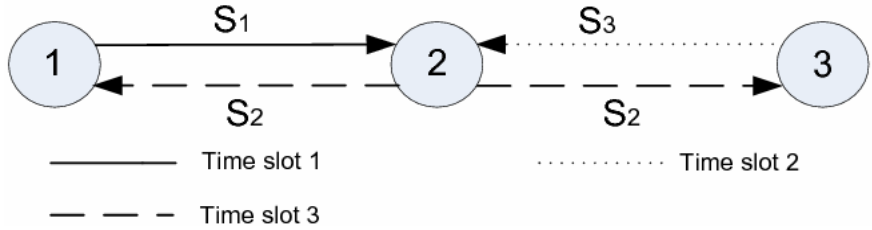

Figure 3. Straightforward network coding scheme

We now introduce PNC. For the time being, let us also assume perfect symbol-level time, carrier synchronization (we will remove this assumption in later sections), and the use of power control, so that the frames from $N_1$ and $N_3$ arrive at $N_2$ with the same phase and amplitude. The combined passband signal received by $N_2$ during one symbol period is

$$\begin{aligned} r_2(t) &= s_1(t) + s_3(t) \\ &= [a_1 \cos(\omega t) + b_1 \sin(\omega t)] + [a_3 \cos(\omega t) + b_3 \sin(\omega t)] \\ &= (a_1 + a_3)\cos(\omega t) + (b_1 + b_3)\sin(\omega t) \end{aligned} \quad (3)$$

where $s_i(t)$, $i = 1$ or 3, is the passband signal transmitted by $N_i$ and $r_2(t)$ is the passband signal received by $N_2$ during one symbol period, $a_i$ and $b_i$ are the QPSK modulated information bits (in-phase and quadrature-phase respectively) of $N_i$; and $\omega$ is the carrier frequency. Then, $N_2$ will receive two baseband signals, in-phase ($I$) and quadrature phase ($Q$), respectively, as follows:

$$\begin{aligned} r^I &= a_1 + a_3 \\ r^Q &= b_1 + b_3 \end{aligned} \quad (4)$$

Note that $N_2$ cannot extract the individual information transmitted by $N_1$ and $N_3$, i.e., $a_1$, $b_1$, $a_3$ and $b_3$, from its received combined signal $r^I$ and $r^Q$. However, $N_2$ is just a relay node and it does not care what the information is. As long as $N_2$ can transmit the necessary information to $N_1$ and $N_3$ for extraction of $a_1$, $b_1$, $a_3$, $b_3$ over there, the end-to-end delivery of information will be successful. For this, all we need is a special modulation/demodulation mapping scheme, referred to as *PNC mapping* in this paper, to obtain the equivalence of GF(2) summation of bits from $N_1$ and $N_3$ at the physical layer.

Table 1 illustrates the idea of PNC mapping. In Table 1, $s_j^{(I)} \in \{0, 1\}$ is a variable representing the in-phase data bit of $N_j$ and $a_j \in \{-1, 1\}$ is a variable representing the binary modulated bit of $s_j^{(I)}$ such that $a_j = 2s_j^{(I)} - 1$. A similar table (not shown here) can also be constructed for the quadrature-phase data by letting $s_j^{(Q)} \in \{0, 1\}$ be the quadrature data bit of $N_j$, and $b_j \in \{-1, 1\}$ be the binary modulated bit of $s_j^{(Q)}$ such that $b_j = 2s_j^{(Q)} - 1$.

Table 1. PNC Mapping: modulation mapping at $N_1$, $N_2$; demodulation and modulation mappings at $N_3$

| Modulation mapping at $N_1$ and $N_3$ | | | | Demodulation mapping at $N_2$ | | |
|---|---|---|---|---|---|---|
| Input | | Output | | | Modulation mapping at $N_2$ | |
| | | | | Input | Input | Output |
| $s_1^{(I)}$ | $s_3^{(I)}$ | $a_1$ | $a_3$ | $a_1 + a_3$ | $s_2^{(I)}$ | $a_2$ |
| 1 | 1 | 1 | 1 | 2 | 0 | -1 |
| 0 | 1 | -1 | 1 | 0 | 1 | 1 |
| 1 | 0 | 1 | -1 | 0 | 1 | 1 |
| 0 | 0 | -1 | -1 | -2 | 0 | -1 |

With reference to Table 1, $N_2$ obtains the data bits:

$$s_2^{(I)} = s_1^{(I)} \oplus s_3^{(I)}; \qquad s_2^{(Q)} = s_1^{(Q)} \oplus s_3^{(Q)} \tag{5}$$

It then transmits, according to the QPSK modulation mapping,

$$s_2(t) = a_2 \cos(\omega t) + b_2 \sin(\omega t) \qquad (6)$$

Upon receiving $s_2(t)$, $N_1$ and $N_3$ can derive $s_2^{(I)}$ and $s_2^{(Q)}$ by ordinary QPSK demodulation. The successively derived $s_2^{(I)}$ and $s_2^{(Q)}$ bits within a time slot will then be used to form the frame $S_2$. In other words, the operation $S_2 = S_1 \oplus S_3$ in straightforward network coding can now be realized through PNC mapping.

As illustrated in Fig. 2, PNC requires only two time slots for the exchange of one frame (as opposed to three time slots in straightforward network coding).

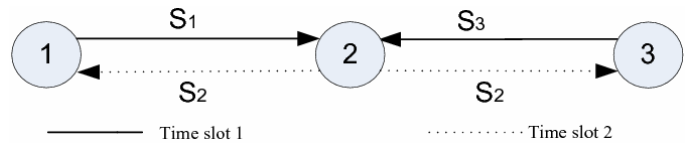

Figure 4. Physical layer network coding

In [4], we analyze the bit error rate (BER) of $S_1 \oplus S_3$ at the relay node for PNC scheme and the straightforward network coding scheme. It is shown that PNC slightly outperforms the straightforward network coding scheme, and slightly underperforms the standard point-to-point BPSK transmission. When the per-hop BER is low, the end-to-end BER for the three schemes is very similar. The main advantage of PNC, however, is the reduced number of time slots needed. In this paper, we showed that this conclusion is true even without perfect synchronization.

### III. PERFORMANCE PENALTY OF SYNCHRONIZATION ERRORS

The basic PNC scheme presented thus far requires symbol-time, carrier-frequency and carrier-phase synchronizations, although these requirements could be relaxed for other variations of PNC [17, 18]. We now consider the performance penalty of synchronization errors on PNC[1]. This framework is applicable to situations where synchronization is not perfect (e.g., synchronization may become imperfect with time due the change of channel) as well as where synchronization is not performed at all. The discussion here is based on the 3-node model in section II.

*1. Penalty of carrier-frequency/phase synchronization errors:*

---

[1] Our paper tries to show the advantages of PNC over traditional and straightforward schemes even with synchronization errors. Although the power synchronization error also affects the PNC performance, it will affect the traditional and straightforward schemes similarly. Therefore, we do not analyze the power synchronization error penalty in this paper.

We first consider carrier-phase and carrier-frequency errors. For QPSK modulation, the two received signals from node $N_1$ and $N_3$ can be written as:

$$r_2(t) = [a_1 \cos(\omega t) + b_1 \sin(\omega t)] \\ + [a_3 \cos((\omega + \Delta\omega)t + \Delta\theta) + b_3 \sin((\omega + \Delta\omega)t + \Delta\theta)] \quad (7)$$

where $\Delta\theta$ is the phase offset and $\Delta\omega$ is frequency offset. Here we assume that the relative carrier-phase offset of the two input signals are known to the receiver[2]. The receiver down-converts the passband signal to the baseband to obtain

$$\begin{aligned} r_2 &= s_1 + s_3 \\ &= a_1 + b_1 + a_3 \cos(\Delta\omega T + \Delta\theta) + b_3 \sin(\Delta\omega T + \Delta\theta) \\ &= [a_1 + b_1] + [a_3 \cos(\theta) + b_3 \sin(\theta)] \end{aligned} \quad (8)$$

where $T$ is the symbol duration and $\theta = \Delta\omega T + \Delta\theta$ is the final phase offset generated by the carrier frequency offset and the carrier phase offset. Hereafter, we only consider the final phase offset $\theta$ without differentiating the contributions of carrier phase and carrier frequency offset. Note that we only need to deal with the case when $-\pi/4 \leq \theta < \pi/4$. If $\theta = \theta' + k \cdot \frac{\pi}{2}$ and $-\pi/4 < \theta' \leq \pi/4$, we can simply replace $a_3$ with $a_3' = a_3 \cdot e^{k\pi/2}$, and replace $\theta$ with $\theta' = \theta - k\pi/2$; $a_3'$ can be mapped back to $a_3$ in a unique manner after detection.

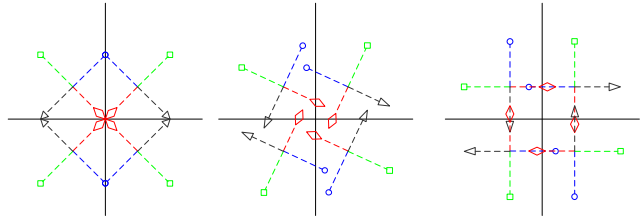

Figure 5. Constellations of superimposed baseband signals in (8) with phase difference $\theta$ equal to $0°, 15°$ and $45°$. There are four possible values for $s_1$ (i.e., $s_1 = \pm 1 \pm j$) as indicated by the intersections of the blue, green, red, and black lines). In each subfigure, we show the four possible values for $r_2$ given each of the $s_1$. The resulting constellation points with the same shape corresponding to the same value of $s_1 \oplus s_3$

---

[2] Before the adjacent transmitters transmit their data concurrently as per PNC, they could first take turns transmitting a preamble in a non-overlapping manner. The receiver can then derive the phase difference from the two preambles. Frequency and time offsets can be similarly determined using preambles. Note that this is different from synchronization, since the transmitters do not adjust their phase, frequency and symbol-time differences thereafter. The receiver simply accepts the synchronization errors the way they are.

From Fig. 5, we can see that as the phase difference $\theta$ increases from 0 to $\pi/4$ (decreases from 0 to $-\pi/4$), a constellation point of $s_1 \oplus s_3$ may break into several points and the distance between the points of different $s_1 \oplus s_3$ may increase or decrease. The BER performance of demodulating $s_1 \oplus s_3$ from the received signal in Fig. 5 is dominated by the minimum distance between constellation points of different $s_1 \oplus s_3$. As shown in Fig. 5, given the phase difference $0 \leq \theta \leq \pi/4$, the minimum square distance, denoted by $d$, between the points of different $s_1 \oplus s_3$ is

$$d^2 = 4(1-\cos\theta)^2 + 4(1-\sin\theta)^2 \tag{9}$$

We now bound the equivalent power penalty caused by the phase difference $\theta$ by benchmark against a reference system. The transmit power of each source in the original system is 2. Consider a reference system in which the transmit power of each source is $P \leq 2$. With perfect phase synchronization, the minimum distance between adjacent points of different $s_1 \oplus s_3$ of the reference system is $2\sqrt{P/2}$. We can tune $P$ such that the minimum distance of the reference system is shortened to the minimum distance of the PNC system with phase difference $\theta$, i.e., $2\sqrt{P/2} = d$. Effectively, the power penalty of the system with phase difference $\theta$ is $P/2$ (note that $P$ is the effective power and 2 is the actual power in the system with phase difference $\theta$). In particular, the BER performance of the PNC system with nonzero $\theta$ is no worse than that of the reference system with zero $\theta$ and with power thus adjusted. This is because decreasing the transmit power in the reference system reduces the distances among all the different constellation points uniformly, while in the original system, the phase difference reduces the distances between some constellation points and enlarges other distances. In other words, the performance loss of the phase difference $\theta$ is upper bounded by a power penalty given as follows:

$$\Delta\gamma(\theta) \leq P/2 = d^2/4 = (1-\cos\theta)^2 + (1-\sin|\theta|)^2 \qquad -\frac{\pi}{4} \leq \theta \leq \frac{\pi}{4} \; . \tag{10}$$

In Fig. 6, we plot the upper bound in (10) with different phase offset. We can see that the power penalty bound is more than 7dB when the phase offset is about $\pm\pi/4$. However, we need not be too pessimistic. When there is no synchronization, a reasonable assumption is that the phase offset is uniformly distributed over $[-\pi/4, \pi/4]$. In this case, the average upper bounded power penalty is

$$\overline{\Delta\gamma(\theta)} = \frac{2}{\pi} \int_{-\pi/4}^{\pi/4} \Delta\gamma(\theta)d\theta \leq \frac{4}{\pi} \int_{0}^{\pi/4} (1-\cos\theta)^2 + (1-\sin\theta)^2 \, d\theta \qquad (11)$$
$$= -3.4\text{dB}$$

That is, even if carrier phase synchronization is not performed at all, the average SNR penalty is upper bounded by 3.4 dB (the simulation in Fig. 11 shows that the average power penalty is no more than 2dB.). To avoid the worst-case penalty and to obtain the average power penalty performance, the transmitters could intentionally change their phases from symbol to symbol using a "phase increment" sequence known to the receivers (or intentionally increase their carrier frequency). If the phase-increment sequences of the two transmitters are not correlated, then certain symbols are received with low error rates and certain symbols are received with high error rates during a data packet transmission. With good FEC coding, the overall packet error rate can be reduced. This essentially translates the power penalty to data-rate penalty.

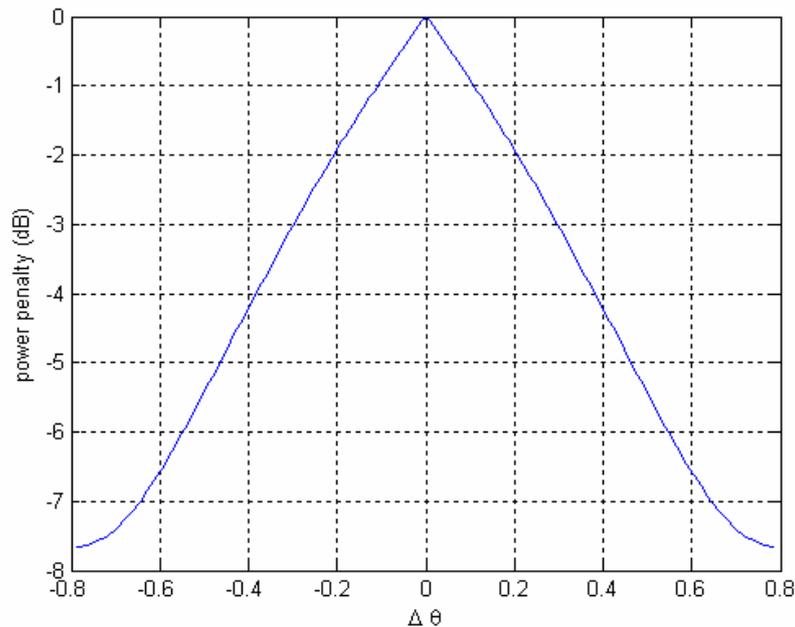

Figure 6. Power penalty upper bound of carrier phase and frequency synchronization errors

Even after taking into account the 3.4dB loss upper bound due to phase asynchrony, PNC may still have a better throughput in the 1-D network and 2-D network than conventional schemes shown in [19]. For 1-D case, the SIR of PNC in [19, eq. (6)] is about 15.3dB, and it is about 11.9dB after subtracting the 3.4dB SIR loss. While the SIR of the traditional 1-D transmission scheme is

$$SIR = \frac{P_0/d^\alpha}{\sum_{l=0}^{\infty} P_0\{2/[(2+4l)d]^\alpha + 1/[(3+4l)d]^\alpha + 1/[(5+4l)d]^\alpha\}} \quad (12)$$
$$= 8.5\text{dB}$$

where $P_0$ is the transmission power, $d$ is the distance between adjacent nodes, and the fading coeffienets $\alpha$ is set to a typical value of 4. For the 2-D case, the SIR of PNC is 13.5dB with $J=5$ is the distance between adjacent PNC chains as in [19, Eq. (9)], and it is 10.1dB after subtracting the 3.4dB loss. It still satisfies the target 10dB SIR requirement as used in traditional wireless networks [19].

*2. Penalty of time synchronization errors:*

Ref. [20] analyzes the impact of time synchronization errors on the performance of cooperative MISO systems, and show that the clock jitters as large as 10% of the bit period actually do not have much negative impact on the BER performance of the system. Based on the similar methodology, we can also analyze the impact of time synchronization error toward the performance of PNC.

In this section, we assume perfect carrier phase and frequency synchronizations for simplicity. In this case, the performance of QPSK is the same as that of BPSK, and therefore we only consider the in-phase signals hence. Let $\Delta t$ be the symbol offset of the two input signals. The two transmitted in-phase signals can be written as:

$$s_1(t) = \sum_{l=-\infty}^{\infty} a_1[l]\cos(2\pi ft)g(t-lT)$$
$$s_3(t) = \sum_{l=-\infty}^{\infty} a_3[l]\cos(2\pi ft)g(t-lT-\Delta t) \quad (13)$$

where, $a_j[l]$ is the $l^{th}$ bit of the real part of signal $s_j(t)$, and $g(t)$ is the pulse shaping signal. The baseband signal can be written as

$$r(t) = r_1(t) + r_3(t)$$
$$= \frac{1}{2}\sum_l a_1[l]g(t-lT) + a_3[l]g(t-lT-\Delta t) \quad (14)$$

After the match filter, the receiver samples the signal at time instances $t = kT - \Delta t/2$ (i.e., at the middle of the offset)[3]. We then have

$$r[k] = r_1[k] + r_3[k]$$
$$= \frac{1}{2}\sum_{l}\{a_1[l]p((k-l)T + \Delta t/2) + a_3[l]p((k-l)T - \Delta t/2)\} \quad (15)$$
$$= (a_1[k] + a_3[k])p(\Delta t/2)/2 + \frac{1}{2}\sum_{l,l\neq k}\{a_1[l]g((k-l)T + \Delta t/2) + a_3[l]p((k-l)T - \Delta t/2)\}$$

where, $p(t)$ is the response of the receiving filter to the input pulse $g(t)$. As widely used in practice, the raised cosine pulse shaping function, $g(t) = p(t) = \frac{\sin(\pi t/T)\cos(\pi \beta t/T)}{\pi t/T \cdot (1 - 4\beta^2 t^2/T^2)}$, is chosen. We see that the time synchronization errors not only decrease the desired signal power, but also introduce inter-symbol interference (ISI). Therefore, we use SINR (signal over noise and interference ratio) penalty here to evaluate the performance degradation. The SINR penalty can be calculated as

$$\Delta\gamma(\Delta t) = SINR(\Delta t) - SNR_0$$
$$= 10\log_{10}(p(\Delta t/2))^2 - 10\log_{10}(\frac{\sigma_{isi}^2 + \sigma_n^2}{\sigma_n^2}) \quad (16)$$

where $\sigma_{isi}^2 = E\{(\sum_{l,l\neq k} a_1[l]p((k-l)T + \Delta t/2) + a_2[l]p((k-l)T - \Delta t/2))^2\}$ is the variance of the inter-symbol interference. Fig. 12 plots the power penalty versus $\Delta t/T$, where the SNR$_0$ is set to 10dB and the roll factor of the raised cosine function is set to 0.5. The worst-case SIR penalty is about -2.2 dB. If we assume the time synchronization error to be uniformly distributed over [-$T$/2, $T$/2][4] (, we can calculate the average SIR penalty as

$$\overline{\Delta\gamma} = \int_{-0.5}^{0.5}\Delta\gamma(\tau)d\tau = \int_{-0.5}^{0.5}SINR(\tau)d\tau - SNR_0 = -1.57dB \quad (17)$$

---
[3] When the symbol time offset is known to the relay node, another choice for the receiver is to discard the inter-symbol interfering part and only take the non-interfering part into account to further improve the performance.

The simulation results in Section V shows that the power penalty due to non-perfect time synchronization is less than 1 dB, which is even smaller than the SINR decrease in (17). In other words, our PNC scheme is more sensitive to the phase offset than the symbol time offset. This fact reminds us that we could adjust the integration time of the match filter at the relay node from T to $T'(T' \leq T)$. Then, the phase offset at the relay is changed from $\theta = \Delta\omega T + \Delta\theta$ to a new value $\theta = \Delta\omega T' + \Delta\theta$. As a result, we may obtain a smaller phase offset (according to the value of $\Delta\omega, \Delta\theta$) at the cost of more time synchronization errors.

---

[4] This assumption is reasonable since one end node can intentionally increase each symbol duration by *T/N* (*N* is the number of symbols in one packet) while the other end node keeps its own symbol duration *T*. As a result, all possible symbol misalignments are experienced at the relay node.

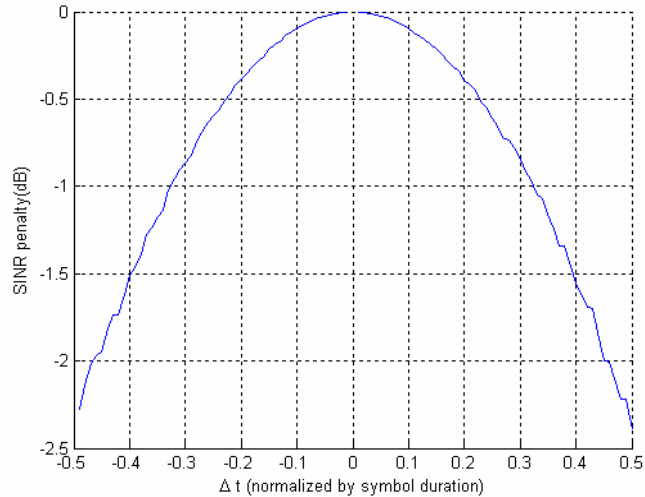

Figure 7. Power penalty of time synchronization errors

Based on the discussion in this section, we can conclude that the performance degradation of 3 to 4dB due to various synchronization errors (including large carrier phase, frequency, and time synchronization errors in the case where a synchronization mechanism is not used at all) is acceptable given the smaller interference of PNC ([19, Eq. (6) and Eq. (9)]) and given the more than 100% throughput improvement obtained by PNC.

IV. SYNCHRONIZATION IN N-NODE PNC CHAIN

In the previous section, we argue that PNC detection is not very sensitive to synchronization errors in the case of $N = 3$. It may appear at first glance that the synchronization problem of the *N*-node case may cause PNC to break down for large *N*, due to the propagation of synchronization errors along the chain. Here, we argue that the detection scheme in PNC does not break down just because *N* is large. In particular, we argue that if the synchronization errors can be bounded in the 3-node case, they can also be bounded in the general *N*-node case.

Synchronization between multiple sources and one destination has been extensively studied in previous works [5-14]. We assume that the feasibility of synchronization in a 3-node chain is a given based on these prior results (i.e., the two end nodes are the sources and the relay is the destination in the set-up of the 3-node chain). Let us consider how the *N*-node case can make use of 3-node synchronization. A possible approach is to partition the long chain into multiple 3-node local groups, as illustrated in Fig. 8, and then synchronize them in a successive manner. Suppose that the synchronization for 3-node can be achieved with reasonable error bounds for phase, frequency, and time (see Section III, where we argue that PNC detection is not very sensitive to synchronization errors), represented by, say, $\theta$, $2\Delta\omega$, $\Delta t$. An issue is the impact of these errors on the $N$-node chain.

For $N$-node synchronization, let us divide the time into two parts: the synchronization phase and the data-transmission phase, as shown in Fig. 9. These two phases are repeated periodically, say once every $T_P$ seconds. The synchronization phase lasts $T_S$ seconds and the data transmission phase lasts $T_D$ seconds, with $T_S + T_D = T_P$. The PNC data transmission described in Section II comes into play only during the data transmission phase. The synchronization overhead is $T_S / T_P$, with $T_S$ depending on the synchronization handshake overhead, and $T_P$ depending on the speed at which the synchronizations drift as time progresses. That is, the faster the drift, the smaller the $T_P$, because one will then need to perform resynchronization more often. It turns out that the $N$-node case increases the $T_S$ required, but not the $1/T_P$ required as compared to the 3-node case, as detailed below.

For the $N$-node chain, let us further divide the synchronization phase into two sub-phases. The first sub-phase is responsible for synchronizing all the odd-numbered nodes and the second for all the even-numbered nodes. We describe only sub-phase 1 here (phase 2 is similar). With reference to Fig. 8, we divide the $N$ nodes into $M = \lfloor (N-1)/2 \rfloor$ basic groups (BGs) and denote them by BG $j$, where $j$ is index of the BGs. Let $\Delta t_{BG}$ be the time needed to synchronizing the two odd nodes in one BG (using, say, one of the prior methods proposed by others). Consider BG1. Let us assume that it is always the case that the right node (in this case, node 3) attempts to synchronize to the left node (in this case, node 1). As an example of the synchronization scheme, node 2 may estimate the frequency difference and phase difference of the signals from node 1 and node 3. Node 2 then forwards the differences to node 3 for it to adjust its own frequency and phase. After this synchronization, the phase, frequency and time errors between nodes 1 and 3 are $\theta$, $2\Delta f$, $\Delta t$. In the next $\Delta t_{BG}$ time, we then synchronize node 5 to node 3 in BG2. So, a total of time of $M\Delta t_{BG}$ are needed in sub-phase 1. Including sub-phase 2, $T_S = (N-2)\Delta t_{BG}$.

It turns out that with a cleverer scheme, sub-phase 2 can be eliminated and $T_S$ can be reduced roughly by half. But that is not the main point we are trying to make here. The main issue is that with the above method, the bounds of the synchronization errors of node $N$ with respect to node 1 become $M\theta$, $2M\Delta\omega$, $M\Delta t$ and these errors grow in an uncontained manner as $N$ increases! In particular, will

PNC therefore break down as *N* increases?

Recall that for PNC detection, a receiver receives signals simultaneously from only the two adjacent nodes. By applying a channel coding scheme [21] at all the nodes, the relay node can recover $S_1 \oplus S_2$ from the received signal without any error. Only the synchronization error penalty within one BG can affect the PNC performance and the penalty will not propagate to other BGs. For example, say, *N* is odd. The reception at node 2 depends only on the synchronization between nodes 1 and 3; and the reception at node *N*-1 only depends on the synchronization of nodes *N*-2 and *N*. In particular, it is immaterial that there is a large synchronization error between nodes 1 and *N*. So, the fact that the end-to-end synchronization errors have grown to $M\theta$, $2M\Delta\omega$, $M\Delta t$ is not important. Only the local synchronization errors, $\theta$, $2\Delta\omega$, $\Delta t$, are important. The same reasoning also leads us to conclude that how often synchronization should be performed (i.e., $1/T_P$) does not increase with *N* either, since it is only the drift within 3 nodes that are important as far as PNC detection is concerned.

Of course, $T_S$ grows with *N*, but only linearly. If $\Delta t_{BG}$ is small compared with $T_P$, this is not a major concern. In practice, however, we may still want to impose a limit on the chain size *N* not just to limit the overhead $T_S$, but also for other practical considerations, such as routing complexities, network management, etc.

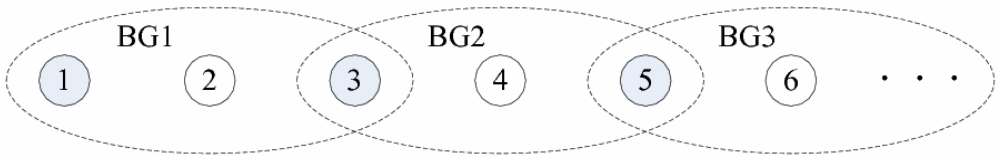

Figure 8. Synchronization for multiple nodes

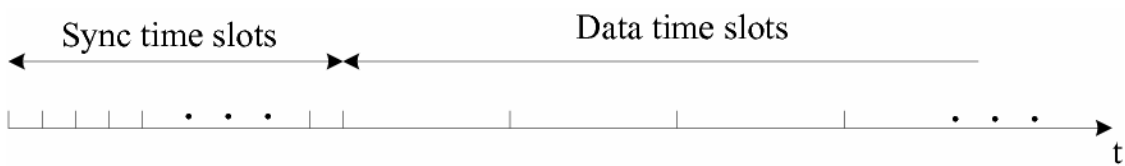

Figure 9. Partitioning of time into synchronization phase and data-transmission phase.

## V. NUMERICAL SIMULATION

In this section, we evaluate the performance loss due to non-perfect synchronization in PNC by numerical simulation. In our simulation, QPSK modulation is used. SNR is defined as the transmission power of each end node over the variance of the Gaussian noise, i.e., $1/\sigma^2$.

We first compare the BER performance with different synchronization levels as in Fig. 10. In this figure, the un-coded BER of $s_1 \oplus s_2$ at the relay node is plotted under different SNRs. The decision rule under perfect synchronization is the same as that in [4], the decision rule for non-perfect phase synchronization is based on ML detection and the decision rule for non-perfect symbol-time synchronization is also the same as that in [4][5]. From this figure, we can see that there is only about 1dB loss when no symbol-time synchronization is not performed, and the performance loss can be ignored when the symbol time offset is randomly distributed in [-0.2$T$, 0.2$T$]. When there are phase synchronization errors, an SNR loss of more than 4dB can be observed at a BER of 3E-3. It is more than the theoretical power penalty of 3.4 dB in (11). The reason is that QPSK is a low order (only 4 constellation points) modulation and it can not efficiently exploit all the signal information when the phase synchronization error is small without the application of channel coding.

---

[5] This decision rule for non-perfect time synchronization is not optimal can be further improved. In that light, the obtained performance in Fig. 9 is only a lower bound.

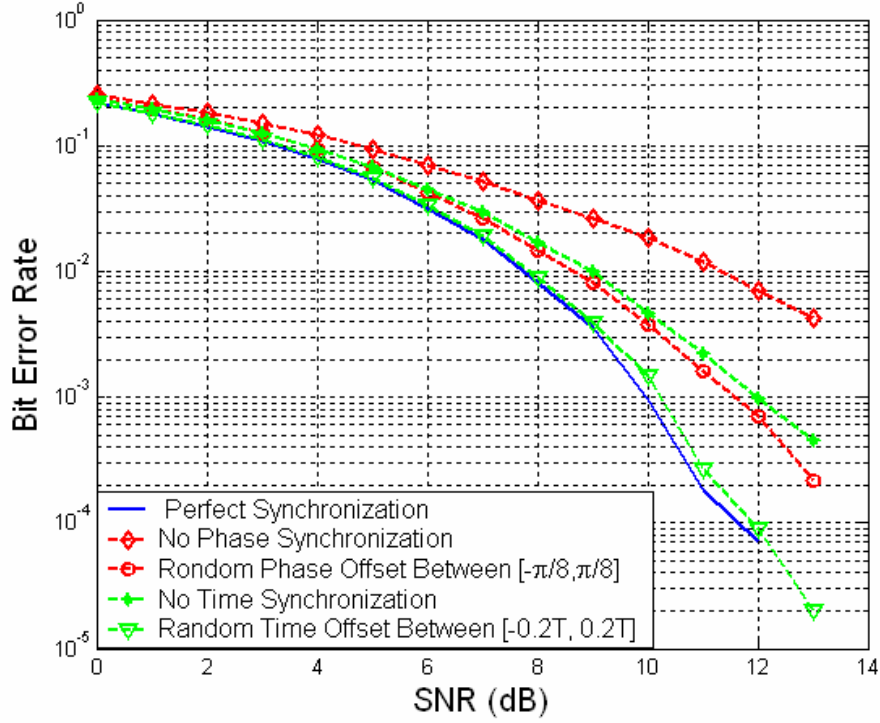

Figure 10. BER performance of PNC with different synchronization levels

We then compare the mutual information performance with different synchronization levels as in Fig. 11. Mutual information is more closely related to the channel coded performance than the uncoded BER. For perfect synchronization, we plot $\frac{1}{2}I(s_1 \oplus s_2; r)$ at the relay node, where the coefficient 1/2 corresponds to the two dimensions in QPSK modulation. For the non-perfect phase synchronization, we plot the mutual information when the phase offset is randomly distributed in $[-\pi/4, \pi/4]$ (this range corresponds to the case of no phase synchronization as mentioned in section III.) as

$$\frac{2}{\pi}\int_{-\pi/4}^{\pi/4} \frac{1}{2}I(s_1 \oplus s_2; r | \theta)d\theta = \frac{2}{\pi}\sum_{k=0}^{19} I(s_1 \oplus s_2; r | \theta = 0.05k\pi/4). \tag{18}$$

For the non-perfect time synchronization error, we simply plot $\frac{1}{2}I(s_1 \oplus s_2; r)$ with time offset randomly distributed in the given range. From Fig. 11, we can see that the SNR loss of no time synchronization is upper bounded by 0.5 dB. This result is much better than the theoretical average SINR loss in (17). The SNR loss of no carrier phase and carrier frequency synchronization is no more than 2dB in the interested SNR region of [0dB, 7dB]. It is also much better than the theoretical upper bound in (11). The simulation results show that the PNC scheme is more robust to synchronization errors than the analysis as in Section III.

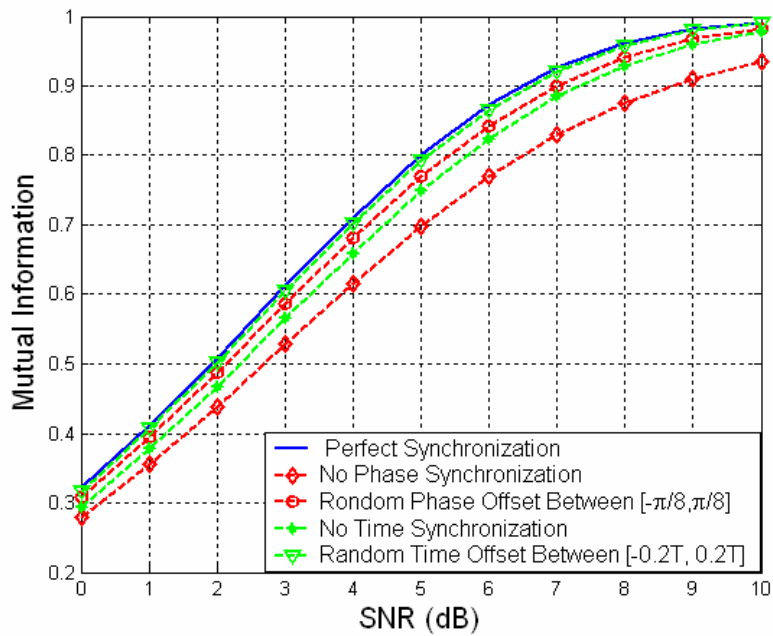

Figure 11. Mutual information performance of PNC with different synchronization levels

## VI. CONCLUSION

This paper investigates the synchronization issues in Physical-layer Network Coding (PNC). We first study the penalty of synchronization errors in PNC. Both analysis and simulation shows that PNC is very robust to the synchronization errors. It has also been shown that the power penalty due to imperfect synchronization can be compensated by the larger SIR in the PNC transmission system. After that, we propose a new synchronization scheme in an *N*-node chain which performs PNC transmission. Last but not least, we have shown that global synchronization in PNC can be achieved without detrimental effects from synchronization-error propagation.